\documentclass[journal,12pt,onecolumn,draftclsnofoot,]{IEEEtran}
\usepackage{cite}
\usepackage{amsmath,amssymb,amsfonts}
\usepackage{algorithmic}
\usepackage{graphicx}
\usepackage{textcomp}
\usepackage{xcolor}
\usepackage{xspace}
\usepackage{url}
\usepackage{subcaption}

\def\BibTeX{{\rm B\kern-.05em{\sc i\kern-.025em b}\kern-.08em
    T\kern-.1667em\lower.7ex\hbox{E}\kern-.125emX}}

\newcommand{\name}{KLASSIFI\xspace}
\newcommand{\acomment}[1]{}
\newcommand{\mcomment}[1]{}

\begin{document}

\title{Knowledge \& Learning-based Adaptable System for Sensitive  Information Identification and Handling }

\author{
\IEEEauthorblockN{Akshar Kaul,}
\IEEEauthorblockA{\textit{IBM Research, India,}
\textit{akshar.kaul@in.ibm.com}}\\

\IEEEauthorblockN{Manish Kesarwani,}
\IEEEauthorblockA{\textit{IBM Research, India,}
\textit{manishkesarwani@in.ibm.com}}\\

\IEEEauthorblockN{Hong Min,}
\IEEEauthorblockA{\textit{IBM T.J. Watson Research Center,}
\textit{hongmin@us.ibm.com}}\\

\IEEEauthorblockN{Qi Zhang,}
\IEEEauthorblockA{\textit{IBM T.J. Watson Research Center,}
\textit{Q.Zhang@ibm.com}}
}

% \author{
% \IEEEauthorblockN{Akshar Kaul, Manish Kesarwani}
% \IEEEauthorblockA{
% \textit{IBM Research, India}\\
% \textit{akshar.kaul@in.ibm.com},
% \textit{manishkesarwani@in.ibm.com}
% }
% \and
% \IEEEauthorblockN{Hong Min, Qi Zhang}
% \IEEEauthorblockA{
% \textit{IBM T.J. Watson Research Center}\\
% \textit{hongmin@us.ibm.com},
% \textit{Q.Zhang@ibm.com}
% }
% }

\maketitle

\begin{abstract}

Diagnostic data such as logs and memory dumps from production systems are often shared with development teams to do root cause analysis of system crashes. Invariably such diagnostic data contains sensitive information and sharing it can lead to data leaks. To handle this problem we present Knowledge and Learning-based Adaptable System for Sensitive  InFormation Identification and Handling (\name) which is an end to end system capable of identifying and redacting sensitive information present in diagnostic data. \name is highly customizable, allowing it to be used for various different business use cases by simply changing the configuration. \name ensures that the output file is useful by retaining the metadata which is used by various debugging tools. Various optimizations have been done to improve the performance of \name. Empirical evaluation of \name shows that it is able to process large files ($128$ GB) in $84$ minutes and its performance scales linearly with varying factors. This points to practicability of \name.

% Protecting sensitive information during data exchange and storage is an important and challenging task.  To serve various privacy protection use cases, we develop a Knowledge and Learning-based Adaptable System for Sensitive  InFormation Identification and handling (\name). \name uses both knowledge-based and learning-based approaches to identify sensitive personal data in documents. To further reduce false positives, \name provides a feedback mechanism for knowledge base refinement. Based on the needs for data exchange, \name applies various anonymization schemes on sensitive data without losing serviceability. \name is efficiently implemented with optimization techniques such that it is suitable for use cases such as protecting sensitive diagnostic data while still ensuring that the data is timely delivered to product service partners. Empirical evaluation of \name shows that it is able to process large files ($128$ GB) in $84$ minutes and its performance scales linearly with varying factors. This points to practicability of \name.

\end{abstract}

%\begin{IEEEkeywords}
%PII, Sensitive Data Redaction, Data Loss Prevention
%\end{IEEEkeywords}

\section{Introduction}

These days enterprises capture vast amount of data about their customers and business processes. 
This data provides competitive advantage to these enterprises and is very valuable.
% This data is very valuable to these enterprises since it provides them with a competitive advantage. 
It contains sensitive information (about the business and customers) and needs to be protected against any inadvertent leakage to anybody (internal or external to enterprise) who is not authorized to see the data. 
Various government regulations such as the Health Insurance Portability and Accountability Act (HIPAA) \cite{hippa}, the General Data Protection Regulation (GDPR) \cite{gdpr} put the onus on the enterprise to protect the customer data. Failing to do so can lead to huge financial penalties.

The data protection has to be looked from various perspectives. 
First of all data should be protected from attackers outside of the enterprise. For this various techniques such as firewalls, VPN (Virtual Private Networks) etc. are used.
The data needs to be protected internally as well. 
This essentially means that all employee of the enterprise should have access to data only on a need-to-know basis.
% Various employees of the enterprise should have access to data only on a need-to-know basis.
For example, a business analyst runs various analytics processes on the data to extract business insights. A data scientist uses the data to train AI models for various tasks such as automated fraud detection. The analyst and the data scientist do not require access to raw data for their work. Instead they require some aggregate statistics over the data. 
Differential Privacy based solutions are used to handle these use cases. These solutions ensure that the analyst and the data scientist can get statistical information about the data without ever getting access to raw data.
% Various techniques have been developed for this use case. The most widely used solution, in this case, is Differential Privacy which ensures that the analyst and the data scientist can get statistical information about the data without ever getting access to raw data.

Enterprises also store a huge number of text documents which contain sensitive information. These text documents are accessed by various employees at different times to do their jobs. 
However not everyone should have access to full document content. 
Enterprises need a solution which allows an employee to access only those parts of the document which they are entitled to. Rest of the document, especially the sensitive information, should be redacted. Various NLP (Natural Language Processing) based solutions are used for building such frameworks.

% The enterprises may also store a huge number of documents containing natural language text which can contain sensitive information. These text documents are accessed by various employees at different times to do their jobs. The enterprises want that employees should only have access to that part of the document which they require for their work. From the rest of the document, sensitive information should be redacted. NLP (Natural Language Processing) based techniques have been used to build frameworks for protecting sensitive documents.

All applications, including those processing sensitive data, run on the IT infrastructure of the enterprise. Inevitably these applications will experience problem and crash. To identify root cause of the crash, diagnostic data such as logs, traces and memory dumps are captured and shared with appropriate teams for debugging.
% This diagnostic data is then shared with appropriate teams for debugging. 
These teams can be in a geographical region different from where the application is running. In many cases, the diagnostic data is shared with third-party software manufacturers as well. The diagnostic data is very likely to contain sensitive information and sharing it can lead to inadvertent leakage of sensitive data.
It is important that all sensitive information is removed from diagnostic data before it is shared for debugging. 

Existing solutions for this problem have primarily looked at extending the programming language to allow developers to specify which memory locations may contain sensitive data. During diagnostic data capture, data from these locations is not stored or is redacted. This is not enough since developers may not mark all the places where the sensitive data may be present. Additionally, it requires an extension to all the programming languages which are used for developing applications. Another big drawback of this approach is that it does not work for already developed and deployed applications.

In this paper, we present \name (Knowledge and Learning-based Adaptable System for Sensitive  Information Identification and Handling), an end-to-end system capable of identifying and redacting sensitive data from diagnostic data, especially memory dumps. \name takes a generic memory dump as input and outputs a memory dump in which all the sensitive information has been redacted. \name ensures that all the meta information required by debuggers such as page headers etc. is kept intact ensuring that the redacted dump is useful.
% This ensures that the redacted dump is useful. 
\name has a built-in Knowledge Base comprising of a comprehensive suite of identifiers which are able to identify large number of sensitive information types. Additionally \name allows a user to augment this Knowledge Base by adding more domain and user specific identifiers. Even a well defined set of identifiers can miss some sensitive data or incorrectly tag some non-sensitive data. \name has a feedback loop which allows a user to provide feedback about such mis-identifications. The feedback provided by user is used in subsequent runs of \name to improve its accuracy.
% This is detailed later on in this section itself
% \name can be configured by each user according to his requirements. It allows customization of identifiers to be used for analysis. 

% One of the major requirement during analysis of diagnostic data for removal of sensitive information is the time required for such analysis. Diagnostic data is very big in size, often reaching several hundred GBs. 
Time taken for analysis and redaction of sensitive data from diagnostic data, often reaching several hundred GBs, is a major factor towards the utility of such systems.
Our customer survey found that the existing systems in place for analysis and redaction of sensitive information take multiple days to process such huge diagnostic data. This is a major bottleneck in the utility of such systems since it delays identifying the root cause of a system crash.
\name is designed to analyze such big diagnostic data quickly.
\name employs various optimization techniques and adaptively adjusts its processing to meet the response time requirements without sacrificing data protection. 

Each enterprise has its own specific requirements from a system which identifies and redacts sensitive data from diagnostic data. Additionally the requirements change for different types of diagnostic data. There is no one size fit all solution. In this spirit \name has following characteristics which allow it to be customized for different use cases.
\begin{itemize}
\item {\bf Document Parsing:} Diagnostic data comes in various formats. Log files contain text which can be read by a human. On the other hand memory dumps contain binary data with embedded text. Parsing this data requires the character set used for encoding. \name allows a user to customize how it reads and parses the data from the source. This allows \name to be adapted for processing different types of files.

\item{\bf Output Usability:} The diagnostic data, especially memory dump, have a pre-defined structure which is leveraged by various tools to help debugging teams in navigating them. \name ensures that the output dump maintains its structure and only the sensitive data is redacted.

\item {\bf Sensitivity Analysis:} 
Sensitive information is domain and context dependent.
Sensitive data present in a medical application is different from that present in a banking application.
Data considered as sensitive changes based on the output file recipient. Output file sent to teams within the organization can have more data than output file sent to third-party.
% Privacy data addressed by HIPPA \cite{hippa} is different from the data that needs to be protected according to the Payment Card Industry Data Security Standards (PCI DSS) \cite{pcidss}. 
\name allows the user to customize what data is considered as sensitive, allowing \name to be adapted for numerous business use cases.

\item {\bf Redaction Techniques:} Technique used for redaction of sensitive data has a bearing on what operations can be done on the output file. If sensitive data is replaced by a fixed string (such as "This data has been redacted") then output file does not allow differentiating between different types of sensitive data. If such differentiation is required (to build some insights from output files) then techniques like hashing, Format Preserving Encryption should be used. \name allows a user to customize how sensitive data is redacted and hence can be used in various different business use cases.

\end{itemize}

The remainder of the paper is organized as follows. The System Architecture and various modes of operations are explained in Section \ref{sec:arch}.
Detailed description of various components of \name is presented in Section \ref{sec:detailed-desc}.
Section \ref{sec:optimizations} details various optimizations done to improve the performance of \name.
In Section \ref{sec:example} a working example of \name is presented. 
Section \ref{sec:exp} presents the empirical evaluation done to measure performance of \name.
Related works are discussed in Section \ref{sec:relwork}. Lastly, the paper is concluded in Section \ref{sec:conclusion}.

\section{System Architecture}
\label{sec:arch}

\begin{figure*}[!t]
  \centering
    \includegraphics[width=\textwidth]{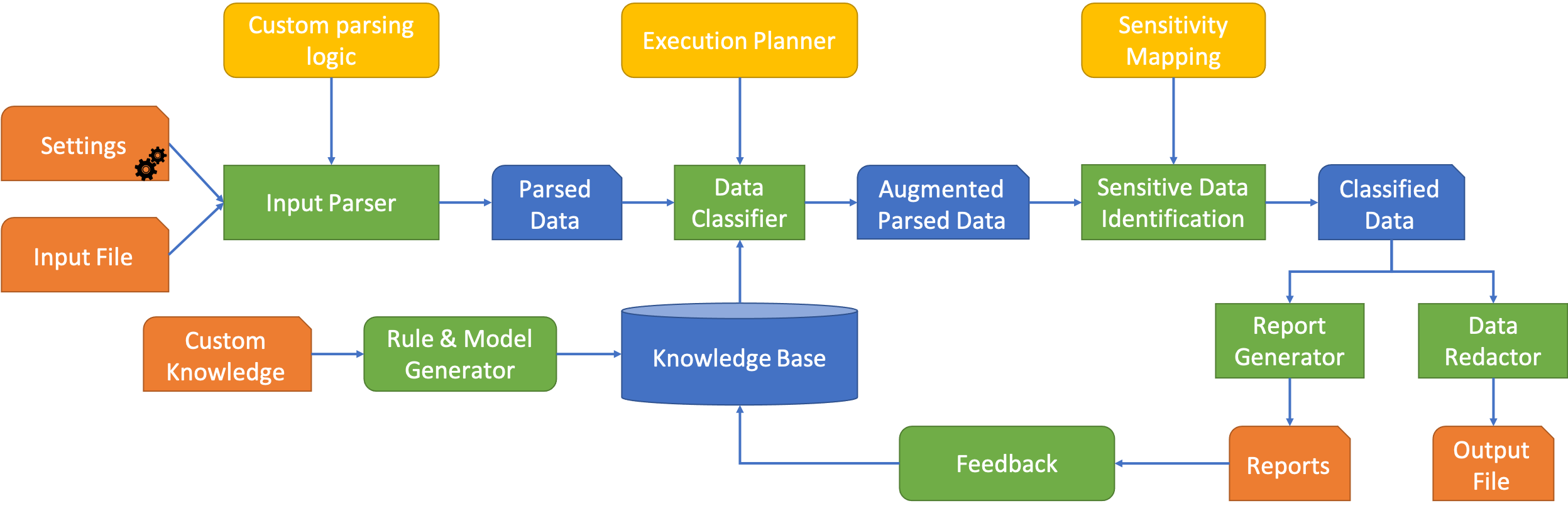} 
  \caption{System Architecture of \name}
  \label{fig:contribution}
\end{figure*}

% In this section, we will present the system architecture of \name and explain its various components. Further,  

In this section, we will present the system architecture of \name and explain its mode of operations. Throughout the paper, we will use the term user, customer and client interchangeably to refer to the same entity -- the consumer of this system.

Figure \ref{fig:contribution} shows the high level architecture of \name. 
Flow of data between various components of \name is indicated by arrow.
% Arrows indicate the flow of data between various components of \name.
\name has following modes of operations:- (a) Analyze (b) Feedback (c) Augment

\subsection{Analyze}
In this mode \name identifies and redacts sensitive data from the input file. This mode takes following inputs:
\begin{itemize}
	\item An input file from which sensitive data has to be identified and redacted.
	\item Settings which allows a user to customize the working of \name according to his specific needs. The configuration is specified by a JSON file which is read by different components of \name to customize their current run.
\end{itemize}
The input file is parsed to get all its data. This data is then analyzed to identify all sensitive information. The sensitive information is redacted to produce the output file. This mode also generates various reports containing information about all the data that was identified as sensitive and non-sensitive.

\subsection{Feedback}
The reports generated during Analyze mode are reviewed by user to check and highlight the data which is mis-classified by \name. The user provides these reviewed reports as input during the Feedback mode. \name uses these highlighted reports to improve its Knowledge Base. This feedback is then used in the subsequent Analyze modes for better accuracy of identifying sensitive information.

\subsection{Augment}
This mode is used for augmenting the Knowledge Base of \name with domain and customer specific information. In this mode user points \name to external files or databases containing sensitive information. \name will add these to the Knowledge Base and later use them for identifying sensitive data in subsequent runs of Analyze mode.

\section{Detailed Description}
\label{sec:detailed-desc}

In this section we will present details about various components of \name and how they work together to make it such a unique system.

\subsection{Input Parser}
The responsibility of \textit{Input Parser} is to parse the input file and create a set of \textit{Parsed Data}. \textit{Parsed Data} is how \name internally represents data and contains enough information for rest of the components to work without requiring knowledge about input file type. This separation allows \name to reuse majority of its components across various file types.

\name has Built-in parser for various commonly used input files including memory dumps taken on mainframes.
In case of a memory dump, the pages belonging to a particular application are scattered. \textit{Input Parser} uses Address Space ids and Logical Addresses to gather these pages and then extracts data for further analysis by using parameters like character set encoding, language etc. It also ensures that the control block information, such as page headers, are not used for further processing by excluding them from \textit{Parsed Data}. This is a very crucial step as it ensures that the output file has all the required meta information used by various debugging tools. Each Built-in parser exposes various configurations, such as character set encoding, system type etc., which can be changed by users according to their use case.
This allows \name to work for a wide variety of use cases by simple configuration changes.

Additionally \textit{Input Parser} exposes a pluggable framework allowing the users to write and plug-in their own custom parsers. This is extremely useful when the file to be analyzed is in a proprietary format. By writing a simple parser customers can reuse rest of the machinery provided by \name.

\subsection{Data Classifier}
\textit{Data Classifier} analyzes the \textit{Parsed Data} and classifies them into various entity types such as Credit Card Number, Social Security Number, Person Name, Address, Email etc. This classification is done with the help of identifiers.
\name supports three types of identifiers :
\begin{itemize}
    \item Dictionary Based Identifiers
    \item Regular Expression Based Identifiers
    \item Machine Learning Based Identifiers
\end{itemize}

\name comes with a rich set of Built-in identifiers. These identifiers constitute its \textit{Knowledge Base}. The \textit{Parsed Data} is passed through each of the identifier in the \textit{Knowledge Base} and all the matches are recorded to create \textit{Augmented Parsed Data}.

Users can enhance the \textit{Knowledge Base} of \name by adding their own custom identifiers and by providing feedback as explained in Section \ref{sec:ingestion} and \ref{ref:feedback} respectively. This allows user to customize and fine tune the classification done by \name according to their specific use case.

This component does bulk of the processing in \name and hence any optimization here has a huge effect on the overall performance of \name. \textit{Execution Planner} module of \name implements various optimizations detailed in Section \ref{sec:optimizations} to improve the performance of \textit{Data Classifier}.

\subsection{Sensitive Data Identification}
This component takes \textit{Augmented Parsed Data} as input and decides which data is sensitive. It is important to note here that all the \textit{Augmented Parsed Data} is not sensitive. Instead \textit{Augmented Parsed Data} contains information about entity types present in data. Which entities are sensitive and which are not depends on the context and use case.

\name uses a mapping between the entity types and sensitivity to decide which \textit{Augmented Parsed Data} is sensitive. \name supports following two types of mappings:
\begin{itemize}
    \item Direct Sensitivity Mapping: This mapping specifies all the entity types that are deemed sensitive by themselves such as Social Security Number. It means that any Social Security Number present in the input file is sensitive and should be redacted.
    
    \item Quasi Sensitivity Mapping: This mapping specifies a group of entities that are deemed sensitive only if all of them are present within a defined vicinity of each other in the input file. For example, a combination of Zipcode and Gender. It means that a Zipcode is considered sensitive only if it is present within the vicinity of a Gender and vice versa. The vicinity is defined as the number of tokens before/after a given token (or a set of memory pages).
\end{itemize}
By default \name treats each entity type it can detect as a Direct Sensitivity Mapping. But it allows customers to provide their own mappings through JSON configuration files. 
% that should be used and also the number of tokens in the vicinity that should be used. 
This user defined mapping allows \name to be customized for various regulations such as HIPPA \cite{hippa}, PCI DSS \cite{pcidss} etc.

The output of this component is classification of Augmented Parsed Data into two sets:- (a) Sensitive (b) Non-Sensitive

\subsection{Data Redactor}
This component generates the \textit{Output File} by redacting the sensitive data identified in the input file. \name supports following techniques for data redaction:-
\begin{enumerate}
	\item {\bf Overwriting Token: } 
	\label{item:pds}
	In this technique, sensitive data is replaced by an overwriting string. \name allows user to provide their desired overwriting string. The overwriting string can either be generic or specific to an entity type. 
	If the length of overwriting string is not equal to the length of the sensitive data then the overwriting string is truncated or replicated so as to make its length equal to the length of sensitive data. 
	
	For example, if overwriting string is ``This data has been redacted" and the sensitive data is ``123 Dummy Street. Seattle, WA 98112'' then the redacted data will be ``This data has been redacted This da''.
	
	\item {\bf Hashing: } 
	\label{item:hash}
	In this technique, sensitive data is replaced by its hash value. \name supports following hashing algorithms:- (a) MD5 (b) SHA-1 and (c) SHA-256.
	If the length of hash value is not equal to the length of the sensitive data then either (a) full hash value is written to output. This implies that the \textit{Output File} length will not be equal to \textit{Input File} length which is fine for certain file types such as system logs. (b) hash value is truncated or replicated so as to make its length equal to the length of the sensitive data. 
	\name allows user to choose the hashing algorithm to be used and how to handle length mismatch.
	
	\item {\bf Encryption: } 
	\label{item:enc}
	In this technique, sensitive data is replaced by its encrypted value. \name supports following encryption schemes: (a) AES (b) FF1 Format Preserving Encryption (FPE).
	AES is an industry standard encryption scheme but it does not ensure that cipher text will be of same length as input. If AES is used then \textit{Output File} length will not be equal to \textit{Input File} length.
	\mcomment{If maintaining file size is a hard requirement, such as for memory dumps, then FPE can be used since it ensures that cipher text is of same length as input.}   
\end{enumerate}

The choice of redaction technique to be used depends on the user requirements. 
If it is required that the \textit{Output File} should have the same size as \textit{Input File}, then Overwriting Token or FPE technique should be used.
If it is required that some parts of the \textit{Output File} be allowed to de-redacted in future then Encryption should be used.
If it is required that de-redaction should not be possible but still it should be possible to get some statistical information from \textit{Output File} then Hashing should be used.

\subsection{Report Generator}
This component generates the following report of the analysis done by \name:-
\begin{itemize}
    \item Sensitive data report containing information about data that has been identified as sensitive and has been redacted from Output File
    \item Non-Sensitive data report containing information about data that has been identified as non-sensitive and is present in Output File in plain text.
\end{itemize}
These reports allow a user to view the analysis done by \name in an easy manner. These reports should have the same access control as the Input File. \name allows encrypting the reports so that only authorized users can view them.
A user can customize \name to generate one or both the reports and also specify whether report should be in plain text or encrypted.

\subsection{Feedback}
\label{ref:feedback}
Even though \name does an excellent job of identifying and redacting sensitive data, sometimes it can mis-identify certain data. This happens mostly in initial phases of deployment when \name does not have much domain specific information. During this time user can manually analyze the reports generated by \textit{Report Generator} and identify such mis-identifications. This is then provided as feedback to \name , which augments its \textit{Knowledge Bas}e with the feedback. This Feedback is then used in subsequent runs by \textit{Data Classifier} to more accurately identify entity types.

\subsection{Rule and Model Generator}
\label{sec:ingestion}
\name comes with a comprehensive suite of identifiers in its \textit{Knowledge Base}. This allows \name to identify a large number of entity types out-of-the-box. But sometimes, this is not enough to detect various domain and business-specific entities. This component allows users to augment \textit{Knowledge Base} of \name by specifying various data sources containing data of these domain-specific entity types. \name supports ingesting data either from an external file or from a database. Additionally, \name allows importing any custom ML model trained by the customer.

\section{Optimizations}
\label{sec:optimizations}
Data Classifier is the main workhorse of \name, which takes the bulk of the processing time. In this section, we detail various optimizations that improve the Data Classifier's performance and hence improve the overall runtime of \name. These optimizations play a big part in bringing \name in the domain of practical and valuable application rather than something that is good to have but is not helpful since it takes days for processing.

% \textcolor{red}{
% Please note, due to space restriction, we have only summarized the ideas in this section. A detailed description of these optimizations, along with implementation details, can be found in the Technical Report \cite{klassifi}.
% }

\subsection{Minimum Identifiers to run}
\label{sec:optimizations-min}

\name comes with a rich set of identifiers in its Knowledge Base, which are used for detecting entity types of data. These Built-in identifiers detect a wide variety of common entity types such as Credit Card Number, Social Security Number, Email etc.
Also, users can augment these Built-in identifiers by adding their own custom identifiers to the Knowledge Base. 

During the \textit{Analyze} mode, the \textit{Data Classifier} component of \name analyzes the \textit{Parsed Data} and classifies them into various entity types. This classification is done by iterating over the list of \textit{Parsed Data} and matching each element against the list of identifiers (Built-in or Custom). This matching of element against an identifier (especially Regular Expression Based Identifier and Machine Learning Based Identifier) is a time consuming process. This processing time is amplified in the setting where the amount of data belonging to the entity types is very low and each matching takes maximum amount of time. 
The number of elements in the list of \textit{Parsed Data} depends on the input file and cannot be reduced. However, \name can reduce the number of identifiers to be used for detecting entity types by utilizing the customer input. This reduction in the number of identifiers to run leads to significant performance improvements.

% Checking each data item against each identifier (Built-in or Custom) is a very time-consuming process. Hence reducing the number of identifiers to run leads to substantial performance improvements.

\name uses the Sensitivity Mapping provided by the customer as input during Analyze Mode to compute the minimal set of identifiers that should be used in the current run. \name starts with an empty list of identifiers. Then it iterates over the user inputs for 
(a) built-in identifiers to be used 
(b) custom identifiers to be used 
(c) dependent identifiers to be used. 
Only those identifiers which are part of these three inputs are added to the identifiers list used to detecting entity types of \textit{Parsed Data}.

The rationale behind this optimization is that if the customer has not mapped an identifier (Built-in or Custom) to be sensitive, either as Direct Sensitivity Mapping or Quasi Sensitivity Mapping, then even if the \textit{Data Classifier} component of \name classifies a Parsed Data as belonging to these entity types and adds them to list of \textit{Augmented Parsed Data}, the \textit{Sensitive Data Identification} component of \name will classify these as Non-Sensitive. As a result these will not be redacted from the output file.
This provides \name with an opportunity to improve its performance by not using these identifiers in the current run. A very important point here is that the output file generated by \name with this optimization is same as the output file generated when this optimization is turned off.

% The rationale behind this optimization is that if the customer has mapped an identifier to be non-sensitive, then even if those entity types are detected, they will not be redacted from the Output File. 
% So \name can improve performance by not using these identifiers in the current run.
% So we may as well not run them. 

The drawback of this optimization is that the Non-Sensitive Report contains entity types for only those identifiers which were part of the minimal set that was used for classifying Parsed Data. The Parsed Data belonging to excluded set of identifiers are shown as not identified in the Non-Sensitive Report. This is a trade-off that enables the customers to ensure that they get the Output File (same as they would get with this optimization turned off) in the shortest possible time.
\name allows the customer to turn off this optimization and get a more detailed Non-Sensitive Report where all the detectable entity types are present. This is usually done as a second pass over the dump. In the first pass, this optimization is turned on to get the Output file in shortest possible time so that it can be used for debugging. In the second pass, when running time is no longer a priority, this optimization is turned off to get the detailed Non-Sensitive Report which can be used for data analysis.

% The drawback of this optimization is that the Non-Sensitive Report now does not contain entity types for such ignored identifiers. But this is a trade-off that enables the customers to ensure that they get the Output File in the shortest possible time. 
% \name allows the customer to turn off this optimization and get a more detailed Non Sensitive Report. This is usually done as a second pass over the dump (for data analysis) when running time is no longer a priority.

\subsection{Minimum Identifiers per vicinity}
\label{sec:optimizations-min-vic}

The identifiers which are part of Quasi Sensitivity Mapping provide further opportunity for reducing the set of identifiers that are used for classifying Parsed Data into various entity types. A set of identifiers which are part of a Quasi Sensitivity Mapping are considered to be sensitive only if all of them appear within the same vicinity. This all or nothing property allows \name to further optimize the set of identifiers that are run in each vicinity. 
\name keeps track of all the identifiers that have till now matched some Parsed Data in a vicinity. If there is no match for an identifier in a vicinity, then all the other identifiers which are part of Quasi Sensitivity Mappings containing this identifier can be skipped for the current vicinity. 

This skipping of identifiers for each vicinity allows \name to reduce the number of identifiers that are run giving a big boost to its performance. This performance boost is amplified even more when a large number of Quasi Sensitivity Mappings are being used. 
A very important point here is that the output file generated by \name with this optimization is same as the output file generated when this optimization is turned off.

Proving that this optimization results in correct Output file is very straight forward. Consider a Quasi Sensitive Mapping whose one of the identifiers has not been found in a vicinity. Now even if \name was to find match for all the other identifiers in the same vicinity, the \textit{Sensitive Data Identification} component of \name will mark them as Non-Sensitive. As a result these will not be redacted from the output file. This leads to the same Output file regardless of whether these identifiers were run for the vicinity or not.

The drawback of this optimization is that the Non-Sensitive Report does not contain entity types for the Parsed Data whose matching with their actual identifier was skipped. 
This is a trade-off that enables the customers to ensure that they get the Output File (same as they would get with this optimization turned off) in the shortest possible time.
\name allows the customer to turn off this optimization and get a more complete Non-Sensitive Report. This is usually done as a second pass over the dump. In the first pass, this optimization is turned on to get the Output file in shortest possible time so that it can be used for debugging. In the second pass, when running time is no longer a priority, this optimization is turned off to get the detailed Non-Sensitive Report which can be used for data analysis.

% Continuing with the optimization of Minimum Identifiers to run, more optimizations can be done for identifiers, which are part of Quasi Sensitivity Mapping. These identifiers are sensitive only if they appear within the same vicinity of other identifiers in the mapping. \name keeps track of identifier matches in each vicinity. If there is no match for an identifier in a vicinity, then all the other identifiers which are part of Quasi Sensitivity Mapping containing this identifier can be skipped. This is because even if a match is found for all the other identifiers, they won't be classified as sensitive. This early exit helps in ensuring that the minimum number of identifiers are run in each vicinity and hence gives a big boost to performance, especially when a large number of Quasi Sensitivity Mappings are being used.

\mcomment{
More details of this optimization can be found in Technical Report \cite{klassifi}. }

\subsection{Dynamic Order of Identifier Evaluation}

The optimizations described till now in Section \ref{sec:optimizations-min} and Section \ref{sec:optimizations-min-vic} focused on minimizing the number of identifiers that should be run. 
Assuming that the list of identifier that should be run is known, the order in which the identifiers are run also has a great effect on the performance of the Data Classifier. \name iterates over the list of identifiers one by one and tries to match the current identifier with the current Parsed Data. If the match is found, the Parsed Data is added to Augmented Parsed Data with the information about the identifier. Further processing of the current Parsed Data is skipped and next Parsed Data in the list is processed. It is easy to see that the for improving performance, the identifier which matches the Parsed Data should be run as early as possible. 

\name leverages the \emph{locality of reference} property of the data contained in the input file to predict which identifiers are more likely to match the current Parsed Data and run them first. The \emph{locality of reference} property specifies that the input files usually have similar data together. 
For example, suppose a SQL select query was fetching data from a relational table when the system experienced failure and a memory dump was taken. In the memory dump, data retrieved by the SQL query will be in close proximity. Additionally, all this data will be of limited entity types since all rows returned by the query will have the same type of data.

\name leverages this \emph{locality of reference} property to change the order of identifiers based on data matches.
\name maintains a sorted list of identifier to be run. Initially this list can be in any order (\name, by default, will sort the identifiers alphabetically). During processing \name maintains this list in Most Recently Used order (Identifier which was most recently matched with a Parsed Data is the first one in the list). 
Whenever a Parsed Data matches an identifier, it is moved to the head of the sorted list. For the next Parsed Data this identifier will be the first one to run. 
This simple algorithm ensures that the identifiers that have a positive match with Parsed Data recently get higher priority than identifiers that don't have a positive match recently.
Additionally this algorithm is quickly able to adapt to the changing \emph{locality of reference} within the input file.
Also this algorithm is fast enough to not become the bottleneck during the analysis.

This optimization allows \name to find the entity type of each Parsed Data much early than an approach with fixed order of identifier evaluation giving a big boost to the performance of the Data Classifier.

\subsection{Processing Modes}
Input Parser divides the Parsed Data into a set of logically dependent objects. For example, If the input file is a text document, Parsed Data is grouped based on paragraphs. If the input file is a memory dump, then Parsed Data is grouped based on the logical address of the page. \name has the following two modes for processing such sets (user can choose which processing mode should be used):
\begin{itemize}
    \item {\bf Concise Mode} 
    \newline
    This is the default mode of processing in \name. In this mode, all the Data in a set is analyzed to find the sensitive data items. Then only these identified sensitive data items are redacted and rest of the data items are copied as it is to output file. This mode ensures that meta-information is left untouched, and hence output file can be analyzed for debugging very easily.
    
    \item{\bf Boolean Mode}
    \newline
    In this mode, \name checks whether a set contains some sensitive data item or not. As soon as the first sensitive data item is found in a set, further analysis of this set is stopped, and an early exit is done. If the set is identified to contain sensitive data, then all the data in the set is redacted in the output file. This mode runs faster than the concise mode due to early exit. However in the output produced by this mode the meta-information present in the Parsed Data list is also redacted. This makes the output file less useful than the one produced by Concise Mode.
\mcomment{but loses meta information in the output file for those sets which contain sensitive information.}
\end{itemize}

Additionally, \name has a dynamic mode in which the user specifies the maximum processing time that \name should take. This mode is helpful when the user is not sure about the amount of sensitive data contained in the input file and cannot decide which mode is best suited for him. 
\name starts processing the input file in Concise mode and estimates the time needed to process the complete input file if execution is continued in this mode. The estimation is based on the time taken to analyze recent pages in the current mode.
If the estimated time is more than the remaining time limit, \name switches to faster Boolean mode for further processing.
\name measures the expected time to completion periodically, and later if the estimated time needed is less than the remaining time, \name switches back to the Concise mode.
If \name estimates that even the Boolean mode will not be able to finish in the required time, it switches to a special mode called skip mode. In this mode, all the data being processed is assumed sensitive and is fully redacted without any classification.
% all the data is fully redacted without any classification.

An important point to note here is that switching to a faster mode allows analysis to finish in the required time but makes the output file less useful. So a trade-off has to be made between the time allocated and the usefulness of the Output File.

% More details of this optimization can be found in Technical Report \cite{klassifi}.

\subsection{Parallelism}
Each set of Parsed Data output from Input Parser is analyzed independently of other sets. This provides an opportunity for parallelism. \name can scale up to any number of available threads, each working on independent sets of Parsed Data. This allows \name to scale up its processing tremendously, and coupled with other optimizations, provides an efficient solution for identifying and redaction of sensitive data from diagnostic dump files.

\section{Example}
\label{sec:example}

% \begin{figure*}[h!]
%   \centering
%     \includegraphics[width=\textwidth,height=\textheight,keepaspectratio]{images/Example_v3.png} 
%   \caption{Example of \name}
%   \label{fig:exam}
% \end{figure*}

\begin{figure*}[h!]
  \centering
    \includegraphics[width=\textwidth]{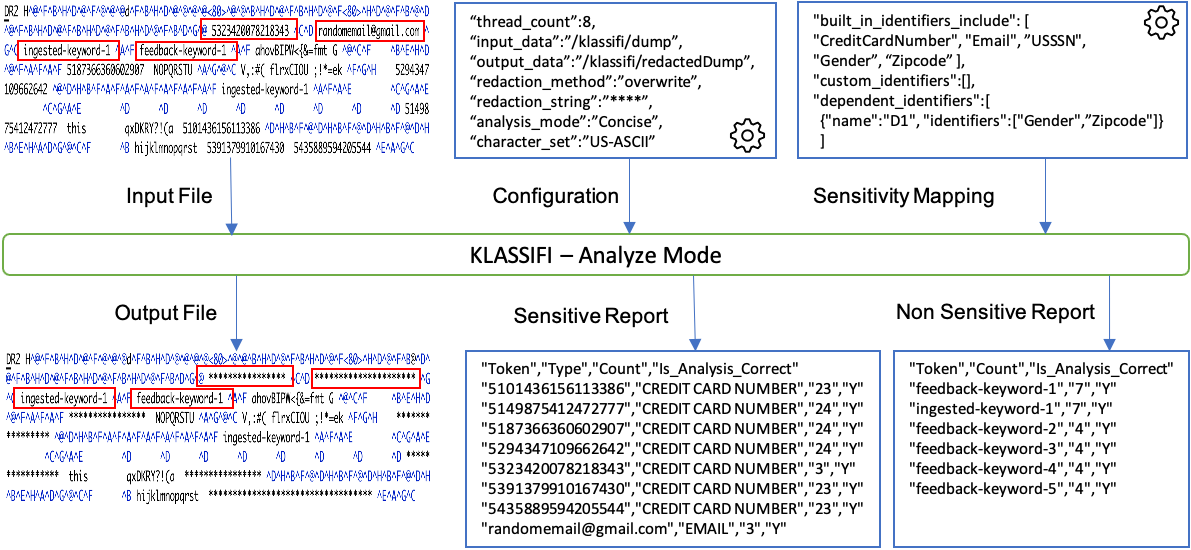} 
  \caption{\name Example - First Analyze Run}
  \label{fig:ex-1}
\end{figure*}

\begin{figure*}[h!]
  \centering
    \includegraphics[width=\linewidth]{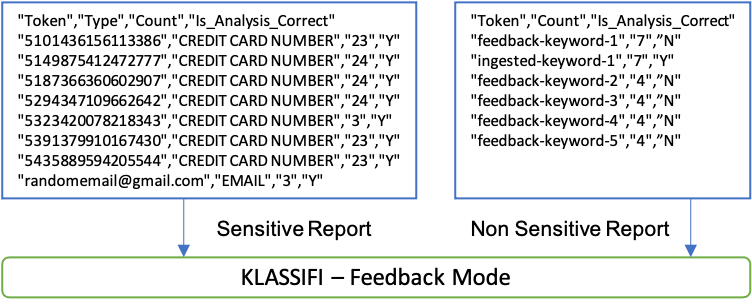} 
  \caption{\name Example - Feedback Run}
	\label{fig:ex-2}
\end{figure*}

\begin{figure*}[h!]
  \centering
    \includegraphics[width=\linewidth]{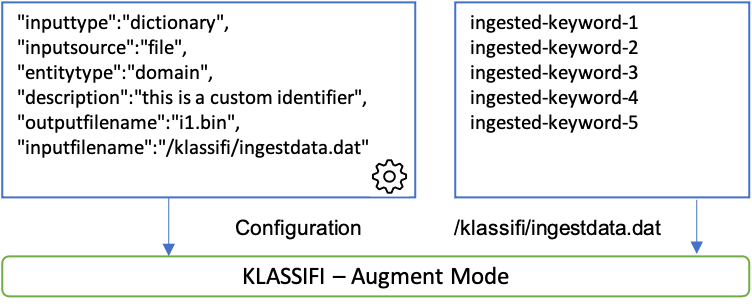} 
	\caption{\name Example - Augment Run}
	\label{fig:ex-3}
\end{figure*}

% \begin{figure*}
% \minipage{0.47\textwidth}
%   \includegraphics[width=\linewidth]{images/ex-2} 
%   \caption{\name Example - Feedback Run}
% 	\label{fig:ex-2}
% \endminipage\hfill
% \minipage{0.47\textwidth}
%   \includegraphics[width=\linewidth]{images/ex-3} 
% 	\caption{\name Example - Augment Run}
% 	\label{fig:ex-3}
% \endminipage
% \end{figure*}

\begin{figure*}[h!]
  \centering
    \includegraphics[width=\textwidth]{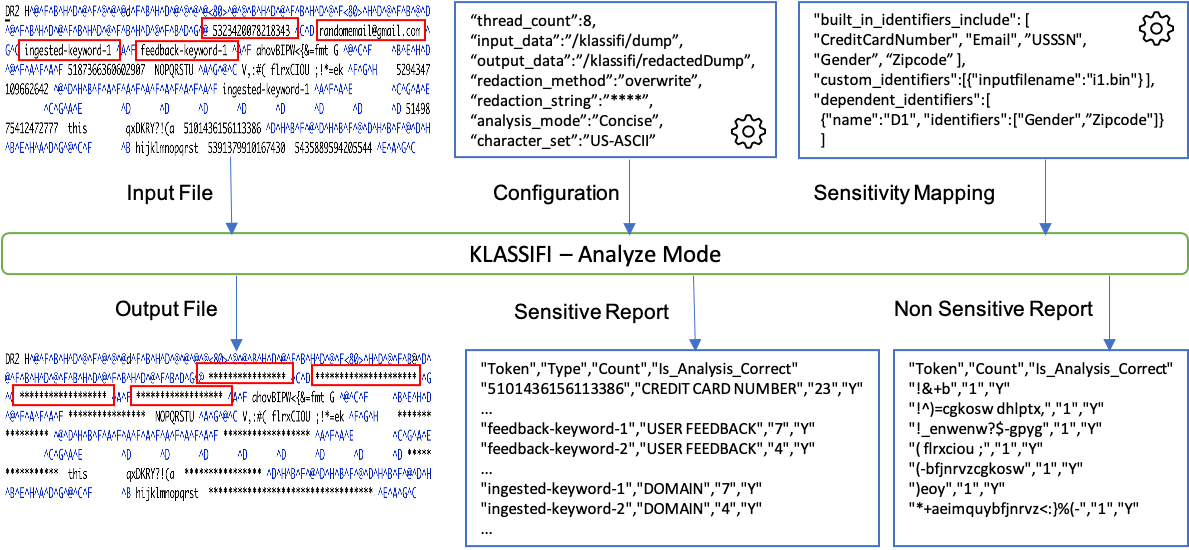} 
  \caption{\name Example - Second Analyze Run}
  \label{fig:ex-4}
\end{figure*}

In this section, we will demonstrate the capability of \name using an example. 
% \textcolor{red}{
% We will only show a subset of information that is useful for understanding the working of \name. 
% More details along with supported configurations can be found in the Technical Report \cite{klassifi}.
% }
The starting point of using \name is when there is an input file from which sensitive data should be redacted.
% To accomplish this \name is run in the Analyze mode with the input file and configuration provided as input.
Figure \ref{fig:ex-1} presents an example of running \name in the Analyze mode. \name takes following inputs:- 
\begin{itemize}
    \item Input File: This example uses a memory dump as an input file. A memory dump is not easy to visualize since it contains hex data. So Figure \ref{fig:ex-1} shows a parsed version of the memory dump for ease of reading. This memory dump contains different types of sensitive data, which includes standard data (Credit Card and Email) and some domain-specific data (\textit{ingested-keyword-1} and \textit{feedback-keyword-1})
    
    \item Configuration: This file specifies various configuration which tunes the current run of \name. This file includes parameters for the maximum number of threads which \name can use, location of the input file and output file, processing mode to be used, redaction method to be used, etc.
    
    \item Sensitivity Mapping: This file provides details about which entity types are considered sensitive. It specifies both Direct Sensitivity Mapping as well as Quasi Sensitivity Mapping.
    
\end{itemize}

Following files are produced as the output of Analyze mode in \name:-
\begin{itemize}
    \item Output File: This is a copy of the Input File from which all data identified as sensitive has been redacted. As can be seen from Figure \ref{fig:ex-1} \name has identified and redacted various sensitive data, including Credit Card Number and Email. However \name is not able to identify domain-specific sensitive data such as \textit{ingested-keyword-1} and \textit{feedback-keyword-1}.
    Also, note that the header information is preserved in the output file without any redaction. This ensures that output file can be used in various standard debugging tools.
    
    \item Sensitive Report: This file contains all the data items that were identified as sensitive by \name. It is a csv file containing information like the entity type and count of how many times this data item was present in the input file. The entity type field makes it easier to understand why the data item was tagged as sensitive.
    
    \item Non Sensitive Report: This file contains all the data items that were not identified as sensitive by \name. It is a csv file containing information about how many times this data item was present in the input file. Note that all the domain-specific sensitive data which \name was not able to identify is contained in this file.
\end{itemize}

The user reviews the reports generated in the Analyze Mode (Sensitive Report and Non-Sensitive Report). If during the review it is found that \name has misclassified some data item, then user marks it by changing the last field (\textit{Is\_Analysis\_Correct}) to \textit{N}. The reviewed and marked reports are used to provide feedback to \name by running it in Feedback mode (Figure \ref{fig:ex-2}). This mode takes a Sensitive Report and a Non-Sensitive Report as input. Only those lines in the reports which have been marked (i.e. \textit{Is\_Analysis\_Correct} is \textit{N}) are used. The rest of the lines are ignored. \name updates its Knowledge Base with the feedback. Specifically, all the marked data items in the Sensitive Report are treated as non-sensitive in the subsequent runs of \name in Analyze mode. Correspondingly all the marked data items in the Non-Sensitive Report are treated as sensitive in the subsequent runs of \name in Analyze mode. The Feedback mode as specified in Figure \ref{fig:ex-2} will update the Knowledge Base of \name to treat \textit{"feedback-keyword-1"} and \textit{"feedback-keyword-2"} as sensitive in subsequent runs of Analyze mode.

\name allows users to augment its Knowledge Base with domain-specific sensitive data. Figure \ref{fig:ex-3} shows an example of this. In this example, an external file containing sensitive data items is being used to augment the Knowledge Base of \name. The configuration file provided as input in this mode contains details of sensitive data and its location. In this example, the sensitive data is of type dictionary (i.e. it is a list of sensitive data items) whose data is present in an external file. The entity type is the name that is shown in the Sensitive Report when a data item of this type is detected. The output file name is where a concise representation of this new type is stored. 
We will like to point here that \name can augment its Knowledge Base with a diverse set of information as detailed in Section \ref{sec:detailed-desc}.

The entity types augmented to the Knowledge Base as described above are not enabled by default in the subsequent runs of \name in Analyze Mode. To enable them, the Sensitivity Mapping has to be updated. Figure \ref{fig:ex-4} shows the second run of Analyze Mode on the same input file after the Feedback mode and Augment mode has been run. 
Note the updated Sensitivity Mapping in this run compared to the first run (Figure \ref{fig:ex-1}). In this run, the augmented entity type is enabled by adding the appropriate information in the custom\_identifier field of Sensitivity Mapping. 
No such configuration change is required for enabling the Feedback given by the user.
As can be seen from Figure \ref{fig:ex-4} \name has now identified and redacted sensitive data which was provided as input using Feedback Mode and Augment Mode.
The Sensitive Report and Non-Sensitive Report also reflect the same.

This example clearly shows how customers can start using \name as a tool that can identify and redact commonly found sensitive data. And then progressively enable it to identify and redact even domain-specific sensitive data by providing appropriate Feedback and by augmenting its Knowledge Base with domain-specific information.

\section{Experimental Evaluation}
\label{sec:exp}

\begin{figure*}[h]
	\centering
	\includegraphics[width=0.75\linewidth]{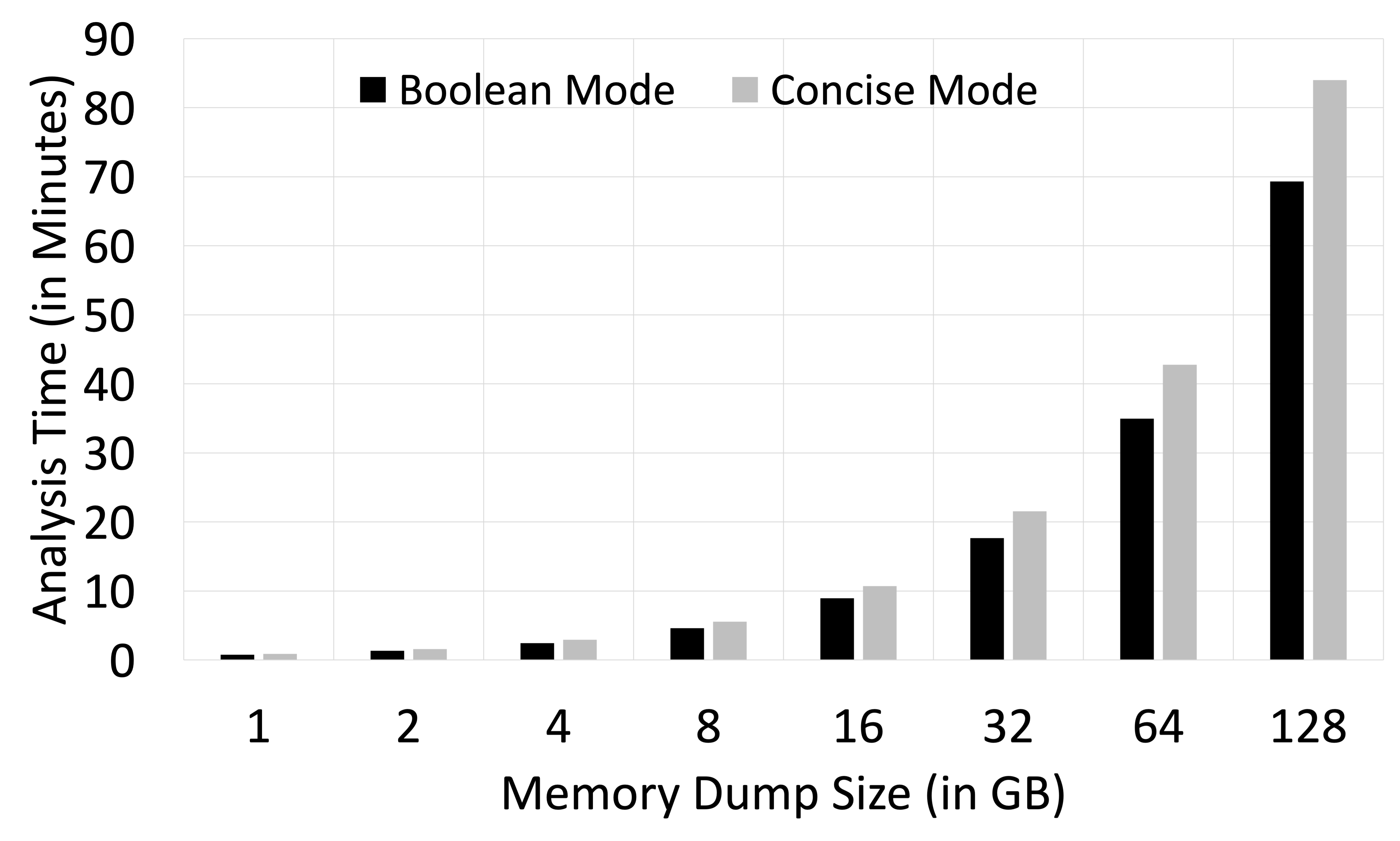} 
	\caption{Analysis time for varying memory dump size (16 threads, 10\% sensitive pages)}
	\label{fig:exp_size}
\end{figure*}

\begin{figure*}[h]
	\centering
	\includegraphics[width=0.75\linewidth]{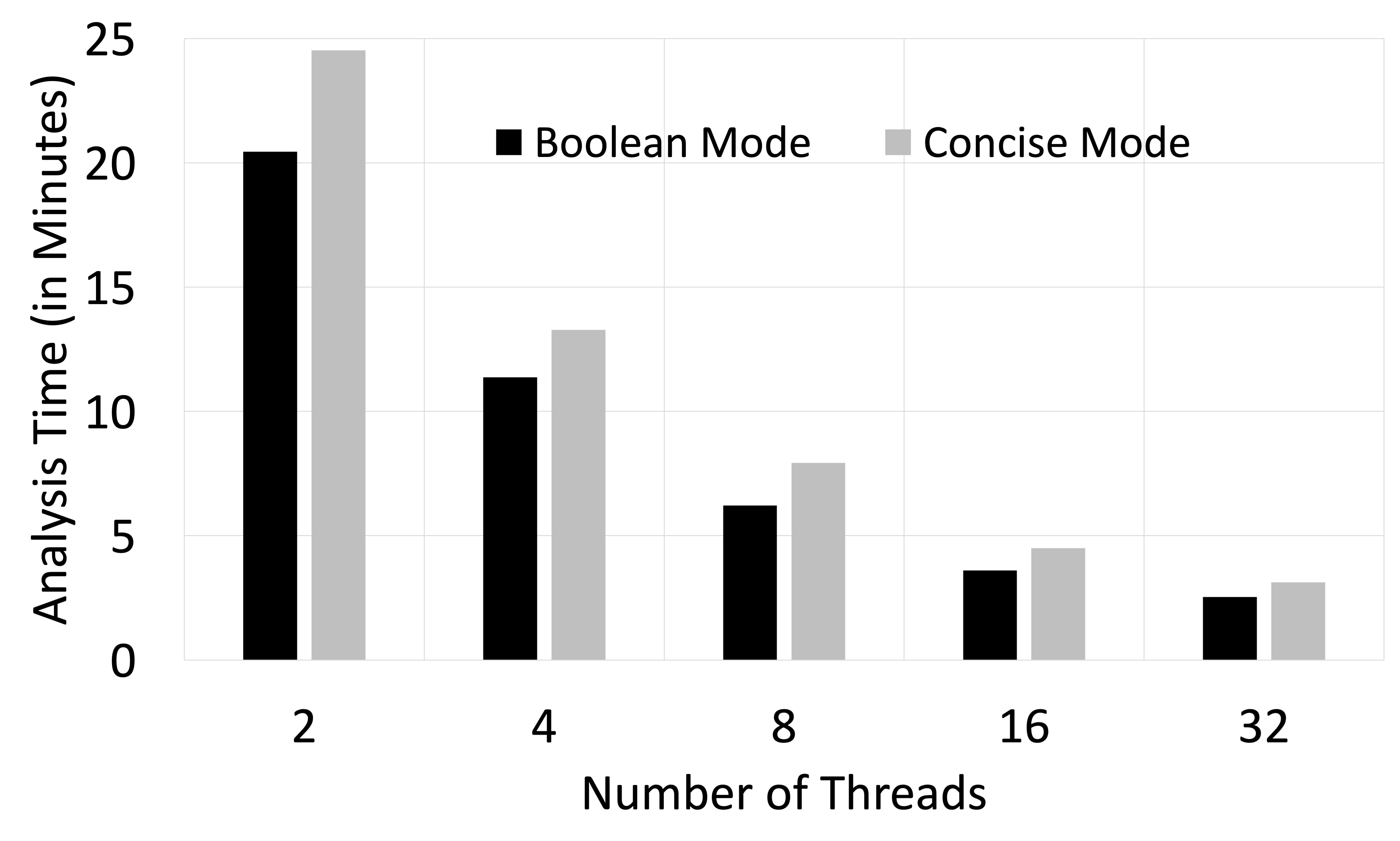} 
	\caption{Analysis time for varying number of threads (8 GB memory dump, 10\% sensitive pages)}
	\label{fig:exp_thread}
\end{figure*}

\begin{figure*}[h]
	\centering
	\includegraphics[width=0.75\linewidth]{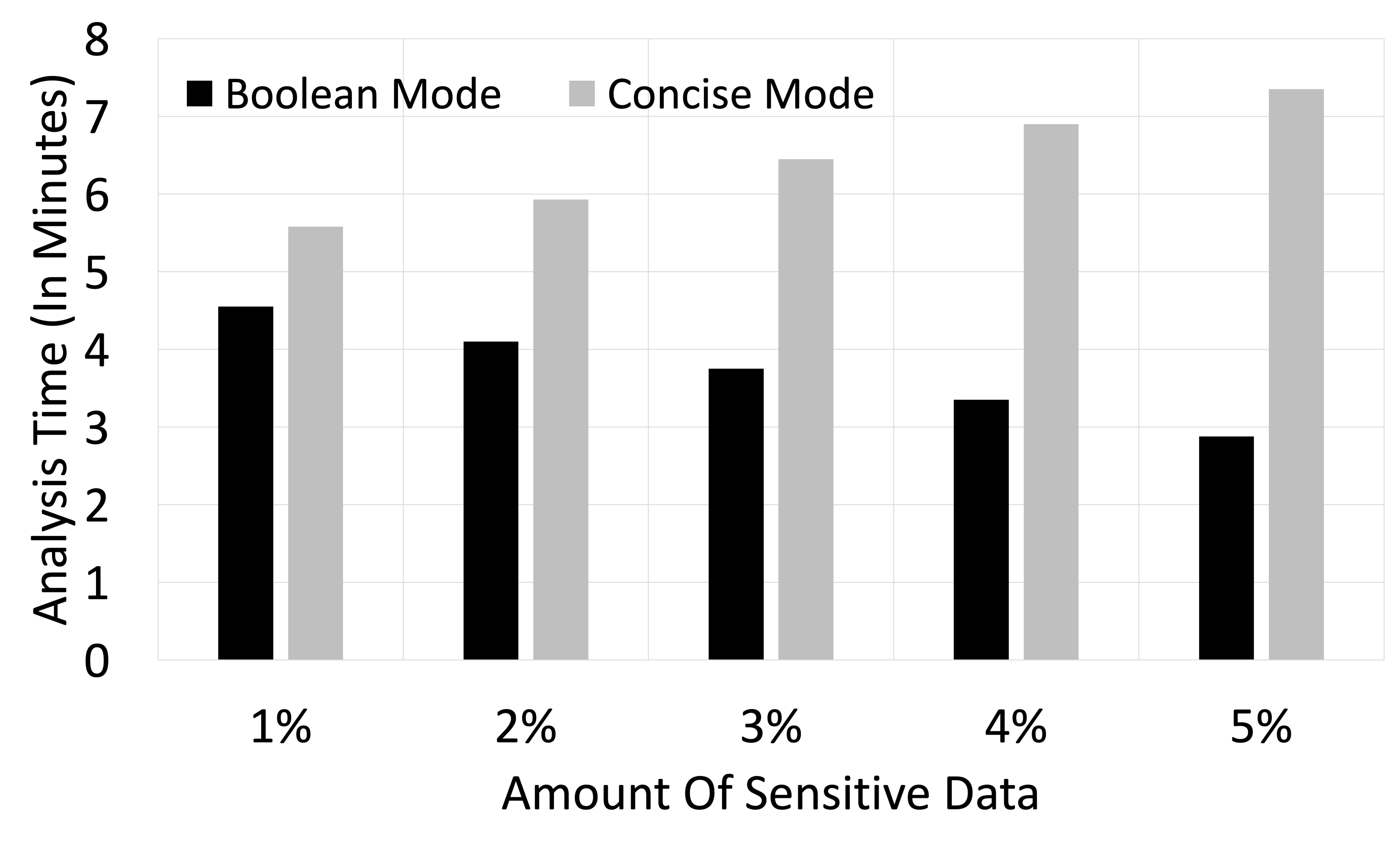} 
	\caption{Analysis time for varying amount of sensitive data (8 GB memory dump, 16 threads)}
	\label{fig:exp_sensitive_data}
\end{figure*}

% Dont show this figure
% \begin{figure}[h]
% 	\centering
% 	\includegraphics[width=\linewidth]{images/exp_varying_sensitive_data_2} 
% 	\vspace{-10mm}    
% 	\caption{Analysis time for varying amount of sensitive data (8 GB memory dump, 16 threads. 10\% sensitive pages)}
% 	\label{fig:exp_sensitive_data_2}
% \end{figure}

\begin{figure*}[h]
	\centering
	\includegraphics[width=0.75\linewidth]{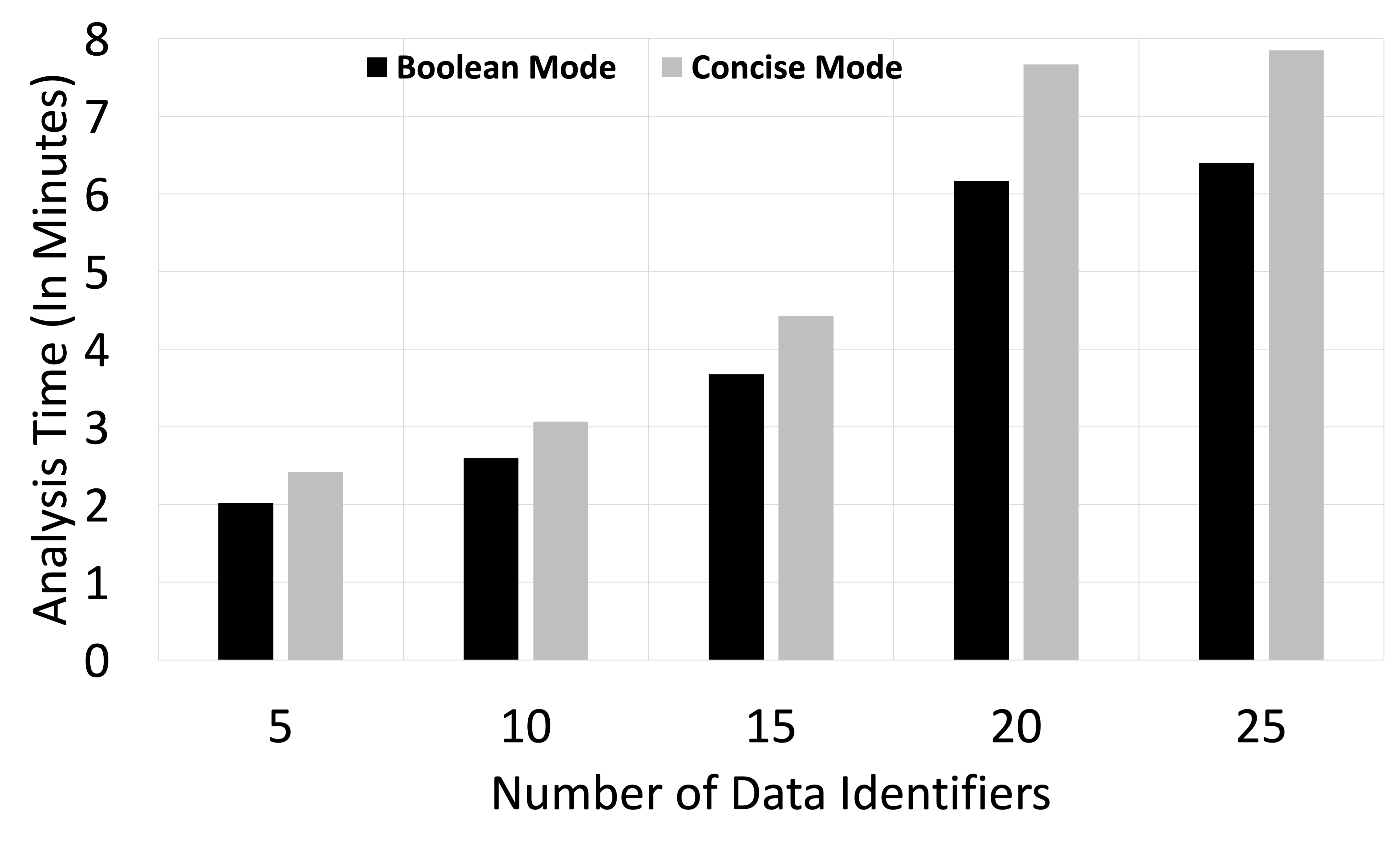}
	\caption{Analysis time for varying Number of data classifiers (8 GB memory dump, 16 threads. 10\% sensitive pages)}
	\label{fig:exp_identifier}
\end{figure*}

\begin{figure*}[h]
	\centering
	\includegraphics[width=0.75\linewidth]{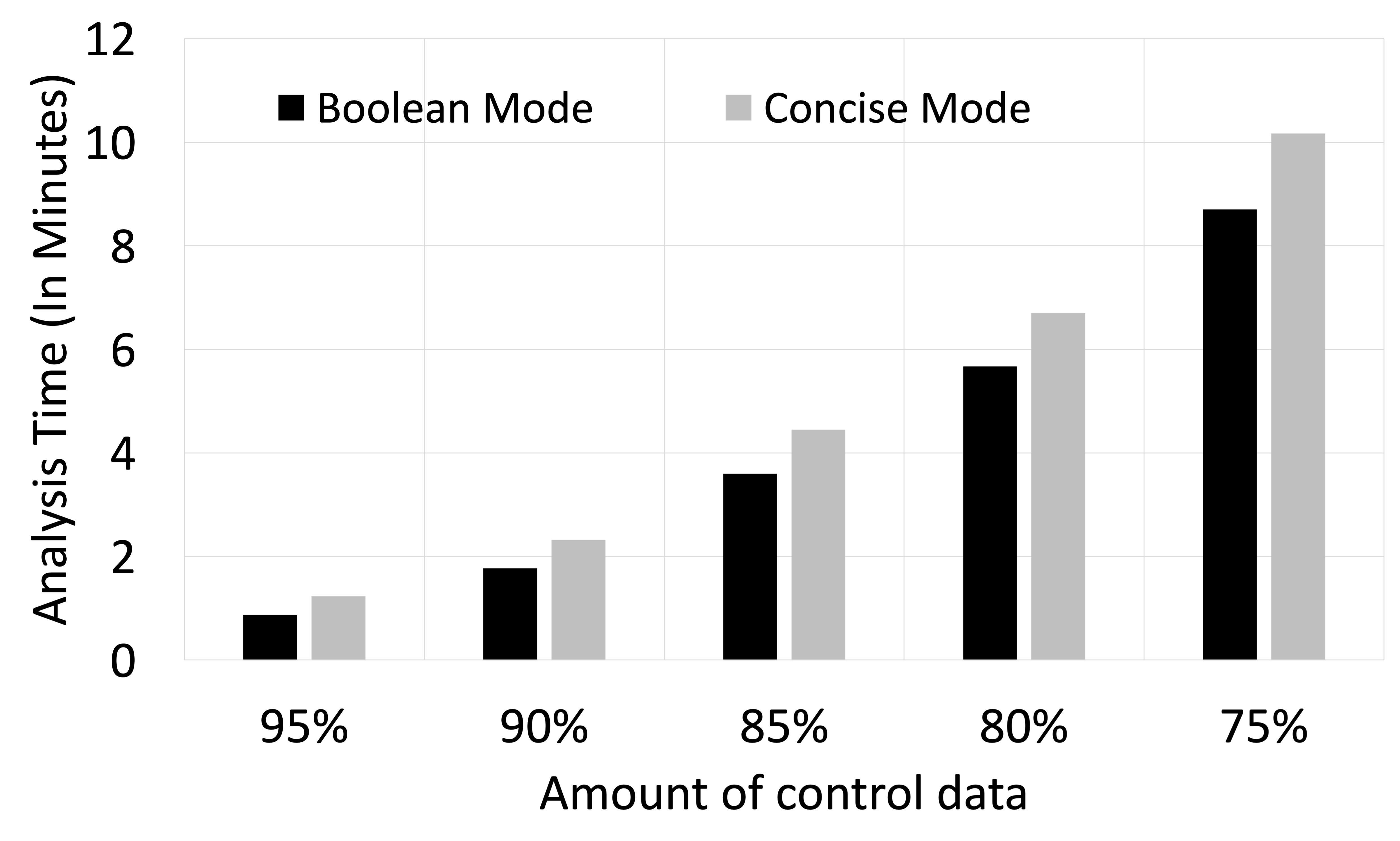} 
	\caption{Analysis time for varying  \%age of control data (8 GB memory dump, 16 threads. 10\% sensitive pages)}
	\label{fig:exp_printable}
\end{figure*}

\begin{figure*}[h]
	\centering
	\includegraphics[width=0.75\linewidth]{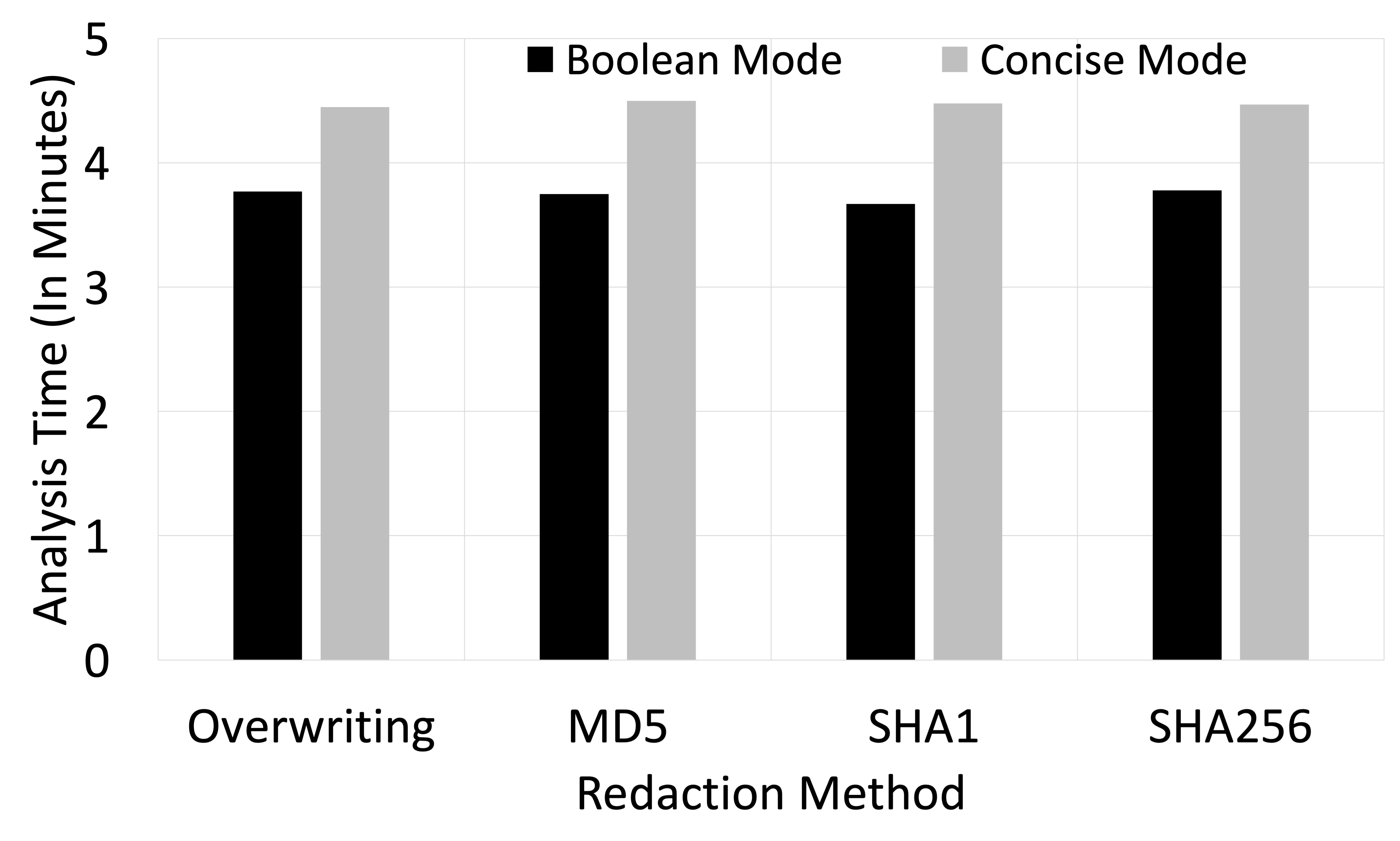} 
	\caption{Analysis time for varying redaction methods (8 GB memory dump, 16 threads. 10\% sensitive pages)}
	\label{fig:exp_redaction}
\end{figure*} 

This section presents the empirical evaluation carried out to measure the performance of \name. To the best of our knowledge, there is no benchmark for evaluating such tools which identify and redact sensitive data from diagnostic data. Using actual diagnostic data such as memory dumps generated by forcibly causing some application to fail is also not a viable option since manually tagging each occurrence of sensitive data, especially in large files, is infeasible.

We did pitch \name against state of the art tool deployed in a real production environment. The system dump captured was first passed through \name and the Output File was given as input to the already deployed tool. The tool was not able to detect any sensitive data in the output given by \name. This points to the fact that \name atleast has the same accuracy as the already deployed tools while providing significantly better performance but in the absence of a benchmark its not possible to quantify it.

Hence we developed a synthetic benchmark of memory dumps by writing a simulated dump generator. This simulated dump generator can generate memory dumps that mimic how an actual memory dump looks like and can control where sensitive data is placed. 
% The synthetic benchmark consists of memory dumps differing in the following aspects:
The simulated dump generator exposes the following configurations while generating a simulated dump:
\begin{itemize}
	\item Size of memory dump
	%\item Size of each page in the memory dump
	\item Percentage of data which is sensitive
	\item Percentage of control data (non-user data part)
	\item Percentage of pages containing sensitive data
\end{itemize}

The running time of \name depends on various factors such as Size of the input file, Number of threads, Percentage of sensitive data, Number of identifiers to run, Percentage of control data \& Redaction technique used. To evaluate the effect of each of these factors, synthetic dumps were generated, which varied individual factors only. The evaluation was carried out on a machine having 32 cores, 128 GB RAM running Ubuntu 18.04. \name is implemented in Java.
Figures \ref{fig:exp_size}$-$\ref{fig:exp_redaction} show the result of this evaluation. 
% The key points from the empirical evaluation are:

\name has been designed from the ground up to analyze really big diagnostic data. Figure \ref{fig:exp_size} shows that running time of \name increases linearly with increasing input size. 
Also, in terms of absolute numbers, \name is able to process a 128 GB input file in about 80 minutes (with most commonly used identifiers enabled and 16 threads). 
This is orders of magnitude faster than existing deployed solutions, which take up to a couple of days to analyze memory dumps of this size (as per our discussions with customers).
This absolute number, along with linear co-relation with input file size, means that \name is able to handle even larger diagnostic data size quickly. This enables dump files to be sent for debugging faster and hence get a faster resolution to the problems in the current production system.
Figure \ref{fig:exp_size} also shows the effect of Boolean mode optimization with Boolean mode always running faster than the Concise mode.

\name lends itself to multi-threading very naturally. Figure \ref{fig:exp_thread} shows how increasing number of threads allows \name to finish faster. This support for parallelism allows customers to provide more resources to \name to ensure that processing of their really big input files completes within the required time.

The amount of sensitive data present in the input file is another big factor affecting the running time of \name. Figure \ref{fig:exp_sensitive_data} shows the running time of \name when the amount of sensitive data per page is constant (1\%), and the number of pages with sensitive data is varied. This corresponds to a situation where sensitive data is distributed across the memory dump. This showcases the effect of Boolean mode optimization. The Boolean mode takes less time as the amount of sensitive data increases (more pages contain sensitive data) as it is able to make an early exit for more pages.

The number of identifiers which \name runs has a huge effect on the running time. This was explored in Section \ref{sec:optimizations} as well. To showcase this, Figure \ref{fig:exp_identifier} shows the running time of \name when the number of identifiers is varied. As is clear from the figure, more identifiers mean more running time. Hence the customer should ensure that only the required identifiers are enabled when running \name by providing proper Sensitivity Mapping.

The amount of control data present in the input file also affects the running time of \name. Input Parser separates out the control data, and only the rest of the data is used for further analysis. Figure \ref{fig:exp_printable} shows that an increase in the amount of control data leads to a decrease in the amount of Parsed Data and hence reducing the running time of \name.

The effect of various redaction methods on analysis time is shown in Figure \ref{fig:exp_redaction}. Redaction methods do not have a noticeable effect on the analysis time (Data Classifier consumes the majority of the time). So a customer can choose a redaction method that suits his business requirement without having to worry about the impact on performance.

All the above experiments point to the efficacy of \name and its applicability to usage in real-world deployments.

\section{Related Work}
\label{sec:relwork}
Sensitive data is ubiquitous in this big data era. For example, a significant amount of sensitive data such as plain text password and email address has been extracted from the crash report of commonly used web browsers \cite{satvat2018crashing}, let alone many other enterprise applications which could contain critical business-related information. As data privacy is becoming an increasing concern, many efforts have been made to minimize the risk of leaking sensitive data, from edge devices like mobile phones \cite{park2016securedom, davis2019systems, hyla2014sensitive, claiborne2020guarding, amiri2017schrodintext} to Cloud data centers \cite{antonatos2018prima, googledlp, macie, shen2018enabling, mahboob2016adopting, ahmadian2018information}, and also with hardware assistants \cite{koning2017no, majzoobi2008testing, eskandarian2019fidelius, kaushik2018secure, baentsch2001protection}.

Ding et al. proposed DESENSITIZATION \cite{ding2desensitization}, which aimed at nullifying the unnecessary data in an application crash dump while keeping the bug- and attack-related data such as the pointers, the heap metadata, and the Return-oriented programming(ROP) gadget chain. In this way, the sensitive data in the crash dump is eliminated while not preventing the third-party vendors from figuring out the cause of the crash.

Broadwell et al. tackled the problem of how to make remote debugging more privacy-preserving \cite{broadwell2003scrash}. The authors designed Scrash, which focused on removing sensitive information from the heap, stack, and global variables. It achieved this goal by introducing customized memory allocation APIs. Users can leverage such APIs to put their sensitive information into a special memory area, which will be wiped out before the crash file is generated.

Similar to Scrash, Feng et al. invented a method \cite{feng2017protecting} to eliminate the sensitive data in a software product by adding a special identifier to the source code of the software, which will be recognized by the compiler and put into a secure data section by the executable file during the runtime. When a core dump is generated, data in the secure section will be considered sensitive and thus eliminated.

Castro et al. \cite{castro2008better} developed an approach to generate the crash dump by using specific input values which is unrelated to the real user input but can be used to reproduce the exact same failure. In this way, the vendors can 
% still 
investigate the execution of the software step by step from the crash dump to detect the bug, but less user-sensitive information is leaked.
% will be leaked.

In addition, to simply eliminate the sensitive information or replace it with random data, kb-Anonymity \cite{budi2011kb} combined k-anonymity model used in the data mining with the concept of program behaviour preservation to redact the sensitive information in a way that sensitive tokens can still be correlated and useful for software testing purposes.

Other than detecting and eliminating the sensitive information in a crash dump, another way to mitigate the security and privacy risk is to optimize the Automatic Crash Reporting System(ACRS) \cite{satvat2020crepe}, which leverages a server to collect crash information from the clients, generate the crash reports to recognize errors that have not been noticed in the development stage. Motivated by the fact that the majority of reports produced by the ACRS server are redundant, CREPE \cite{satvat2020crepe} is proposed to limit the number of duplicated crashes submitted to the server so as the reduce the possible sensitive information leak from the submitted crashes. In CREPE, the client derives a signature of each crash which can be used to query a local datastore to determine whether the detailed dump data needs to be sent to the server or not.

Customer data privacy is critical to the success of many industrial enterprises. Especially with the advances of Cloud computing, in which customer data needs to be stored remotely, it is essential for the Cloud service providers to ensure that the sensitive information in their customer data will not be leaked. To prevent re-identification attacks carried out by exploring personal-specific data, PRIMA introduces \cite{antonatos2018prima}, an end-to-end solution for personal data de-identification, which first identifies privacy vulnerabilities in the datasets, and then performs utility-preserving data masking and data anonymization to eliminate the discovered vulnerabilities.  Yuya et al. \cite{ong2017context} proposed a context-aware DLP(Data Loss Prevention) system, which leveraged machine learning and deep learning techniques to detect real-time sensitive data at different levels, such as documents level, sentence level, and token level. Amazon Macie \cite{macie} has also applied a learning-based approach to protect sensitive data from their customers. Macie uses machine learning models to automatically discover, classify, monitor, and protect sensitive information such as personal identifiable information, protected health information, and financial data in S3 storage. Google Cloud DLP \cite{googledlp} has offered similar services to detect and mask customer-sensitive data, as well as to measure re-identification risk in structured data. Other than these big technology companies, which aim at protecting broad categories of sensitive data, there are also companies focusing on anonymizing a specific type of information. For example, the Synchrogenix ClinGenuity Redaction Management Service (CRMS) \cite{crms} is able to redact sensitive medical records data and claimed a high accuracy.

All of the above techniques are effective in their problem settings. Our \name work addresses sensitive information identification and redaction in diagnostic data from an enterprise environment with new and legacy applications. Using customized memory allocation APIs \cite{broadwell2003scrash}  or tagging the data or memory \cite{feng2017protecting} should be effective for new applications. But it is less practical for legacy applications developed decades ago. A similar practicability issue arises when locating sensitive data in a dump by comparing against a reproduced dump via specified data \cite{castro2008better}.   Nullifying debugging-related information such as pointers  \cite{ding2desensitization}  also has limited applicability in complex software memory dump due to serviceability concerns. Richer contextual information is important in the accuracy of sensitive information identification. Different solutions approach this differently based on their data hosting or processing needs and environments \cite{ong2017context} \cite{googledlp}.   Since dump data, which contains mixed binary and non-binary data, is context weak, \name amends this by learning from the application data, such as database data,  stored within the same system where the dumps are captured, the customization applies to both model and knowledge base that \name uses.  In addition,  \name uses a feedback mechanism to continue to enrich the knowledge and model for improved accuracy. 

\section{Conclusion and Future Work}
\label{sec:conclusion}

In this paper we present \name a Knowledge and Learning-based Adaptable System for Sensitive InFormation Identification and handling. \name is able to identify and redact sensitive data from a wide variety of input files. \name supports a number of customization allowing it to be adapted for a large number of business use cases. We also present various optimizations done to improve the performance of \name and the experimental evaluation showcase the efficacy of \name. 
In future we plan to extend our evaluation and work towards building a standard benchmark for tools like \name.

% In this paper, we propose \name: a Knowledge and Learning-based Adaptable System for Sensitive InFormation Identification and handling. \name accepts structured, semi-structured or unstructured input data file and generates the corresponding anonymised file. The system supports a flurry of optimization techniques to intelligently satisfy stringent user requirements without compromising the security of sensitive data. 
% Our system is fast and scalable compared to the state of the art solutions. The experimental results demonstrate that \name can efficiently identify and anonymise sensitive data for large diagnostic memory dumps within practical time limits ensuring that the data is timely delivered to product service partners and adheres to the required governance requirements. In the future, we plan to extend our evaluation and implement more anonymization approaches like format-preserving encryption and extensively analyse the relational data storage and semi-structured data files.

\bibliographystyle{bibliography/IEEEtran}
\bibliography{bibliography/ref}

\end{document}